PAPER • OPEN ACCESS

# Modular assurance of an Autonomous Ferry using Contract-Based Design and Simulation-based Verification Principles

To cite this article: Jon Arne Glomsrud et al 2024 J. Phys.: Conf. Ser. **2867** 012043

View the article online for updates and enhancements.

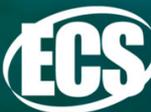

# Modular assurance of an Autonomous Ferry using Contract-Based Design and Simulation-based Verification Principles

**Jon Arne Glomsrud[1]\*, Stephanie Kemna[1], Chanjei Vasanthan[1], Luman Zhao[1], Dag McGeorge[1], Tom Arne Pedersen[1], Tobias Rye Torben[2], Børge Rokseth[3], Dong Trong Nguyen[3]**

[1] DNV AS, Trondheim, Norway
[2] Zeabuz AS, Trondheim, Norway
[3] NTNU, Trondheim, Norway

\*E-mail: jon.arne.glomsrud@dnv.com

**Abstract.** With the introduction of autonomous technology into our society, e.g. autonomous shipping, it is important to assess and assure the safety of autonomous systems in a real-world context. Simulation-based testing is a common approach to attempt to verify performance of autonomous systems, but assurance also requires formal evidence. This paper introduces the Assurance of Digital Assets (ADA) framework, a structured method for the assurance of digital assets, i.e. novel, complex, or intelligent systems enabled by digital technologies, using contract-based design. Results are shown for an autonomous ferry assurance case, focusing on collision avoidance during the ferry's transit. Further, we discuss the role of simulation-based testing in verifying compliance to contract specifications, to build the necessary evidence for an assurance case.

## 1. Introduction

The introduction of autonomous shipping technology presents challenges beyond the technical developments, such as autonomous navigation or docking. Ensuring maritime safety is key, thus it is paramount to assure that autonomous surface vessels (ASVs) operate with at least the same level of safety as human-controlled vessels. An ASV should understand not only operational risk, but also risk stemming from any critical vessel capability. It should not only account for *uncertainties* in sensor measurements, data classification, and predictions, but also *risk* related to manoeuvring capabilities. Only then can an ASV autonomously decide on a safe action.

Consider an autonomous passenger ferry as an example. For this vessel, the basic task is to transport passengers in a safe efficient, and smooth manner, ensuring a comfortable transit for the passengers, in any situation that may arise. This is a considerable challenge due to the complexity of the unstructured environment, the operation, and the autonomous system itself. In addition, safety, efficiency and smoothness are *emergent properties* that arise from the interaction between the vessel and elements of the environment. Furthermore, the ferry's behaviour emerges from interactions amongst its components.

Although there is ongoing research towards end-to-end machine learning for autonomous navigation, e.g. in autonomous driving [1], the majority of cases coming to market currently employ modular approaches. The modular implementations can be seen as consisting of two key components: (1) situational awareness (SITAW) and (2) control. In this work, SITAW is defined as providing the ASV with an understanding of its own states, relevant states of the environment, and an understanding of how these states change over time. SITAW capabilities are primarily defined by the ASV's sensors and sensor processing. For control, the ASV needs to solve a control problem, balancing the task of operating efficiently, safely and smoothly.

Simulation-based testing (SBT) involves testing of software/system components in simulated environments, such as via software-in-the-loop (SIL) testing. Given that simulations can oftentimes be run faster than in real time, and in parallel, SBT is a cost-effective and scalable tool. However, current simulation environments for autonomous system development do not sufficiently provide formal proofs, needed for assurance: Using formal methods, one can provide evidence that a system meets the specified safety requirements. However, formal methods generally do not scale well for complex, autonomous systems [2].

A combination of formal- and simulation-based methods can offer a comprehensive approach to validating the safety of autonomous systems by providing formal evidence, while also scaling well. In this paper, we propose a systematic and modular assurance approach to support the entire lifecycle of an autonomous ferry system. This approach integrates Contract-Based Design (CBD) principles, leveraging formal methods for precise specification and verification of contracts, and complements them with a holistic simulation-based verification strategy to ensure robustness and reliability throughout the system's lifecycle.

*1.1 Related works*
The primary focus of this paper is on the use of CBD for verification and assurance of ASVs. While CBD has previously been used as an aid in system assurance, e.g. [3], and more recently towards modularization and assurance cases, e.g. [4], the use of CBD in the maritime domain is new. In [5], some of this paper's authors introduced the use of CBD for the assurance of ASVs, to present a top-down framework for contract-based assurance. The framework proposes an operational hazard identification to define the top-level component contract and a stepwise refinement of the top-level module into detailed submodules and contracts. Moreover, it proposes the use of a formal theorem prover to verify the correctness of contract refinements and contract compatibility between modules. The framework also incorporates the use of SBT to verify module compliance with contracts. In this paper, we extend [5] by (1) detailing the hazard identification methodology, (2) integrating CBD with assurance case techniques, and (3) further developing the use of simulations in combination with CBD. For a further background into CBD and relevant references, see Section 2.2.

As an alternative approach, [6] looked at methods and regulations used in the automotive industry and proposed a cybersecurity and safety assurance case for ASVs. A *Cybersecurity Assurance Case* was created using goal structured notation (GSN), while applying eliminative argumentation and using the Attack-Defense Tree structure. Their approach used a typical assurance case analysis via; hazard identification, extent of harm, evaluation of risks, mitigation strategies, recording evidence, and evaluating and monitoring. In their start phase, they identified goals, subgoals and context – though it was not specified how these were defined – to then create the attack-defence trees for goals. Their approach is structured, but does not adequately address emergence, and seems to lack a clear system model to connect arguments and evidence with the real systemic risks. In contrast, in this paper, the focus is more on how to set up the assurance case, via an extensive analysis of components, their assumptions and guarantees, to identify requirements and risks, and to lead to evidence needs.

*1.2 Objective & Contributions of this paper*

There remains a research gap in integrating formal methods with SBT methods to address the complexities of assuring ASVs. To bridge this research gap and assure maritime safety, we propose a systematic and modular assurance approach, supporting the entire lifecycle of an autonomous system. The main contribution of this paper is to describe a novel Assurance of Digital Assets (ADA) Framework; a structured method for assurance of digital assets: i.e. novel, complex, or intelligent systems enabled by digital technologies. In addition, we show how utilizing simulations can offer exploration and testing of the contract specifications, to build the necessary evidence for an assurance case, and thus provide the grounds for justified confidence.

## 2. Theory

The modular assurance approach uses CBD principles combined with a holistic simulation-based verification strategy. Key to the success of modular assurance using CBD is accurate and consistent contract specifications of responsibilities, interactions, and dependencies of the system components in the environment they operate. At the component level, assurance requires that a component sufficiently implements its contract, and the contract properly represents its implementation. Similarly, at a system integration level the component interactions and dependencies must match the contract specifications and vice versa. The use of CBD in assurance of maritime control systems is introduced in this paper.

*2.1 Modular assurance using CBD*

The systematic use of contracts in system design is not new, see e.g. [3] [7], though its use is new in the maritime domain. At the core of CBD is the concept of assume-guarantee (A-G) contracts. They specify a set of *assumptions* and a *guarantee*. The guarantee expresses a property that a component should provide. The assumptions are implemented by the environment, i.e. the components that it interacts with and the external world. The component and environment together implement the guarantee. CBD thus provides a formal theory that uses contract refinements to account for dependencies of a component on its environment and subcomponents, to ensure consistency across a complex system. The formal CBD theory facilitates addressing diverse stakeholder concerns, independent development, reuse of components, and modifications to the system and its requirements. This paper uses a practical approach to combine CBD with assurance cases from [7], building on [4].

**Tab. 1** shows an extended CBD contract template developed and used in ADA, to help the system identification (Sec. 2.2), which contains additional, relevant, and practical information to the CBD contracts. This can be used for specifying any component responsibility, within the system, or for the systems or external entities in its environment.

**Table 1.** ADA CBD contract template.

| Field | Description |
| --- | --- |
| Responsibility (R) | The *responsibility* of a component in understandable text. |
| Function (F) | How the *responsibility* is addressed by a *function* and its *outputs*. |
| Input (I) | The *input* interfaces of the component. |
| Assumption (A) | The *assumptions* the component makes about any dependency on conditions, states, events, behaviours, or other component's *guarantees* outside the power of the component itself. |
| Output (O) | The *output* interface of the component. |
| Guarantee (G) | The *guarantees* or promises the component provides about its behaviours under the given set of *assumptions*. |

*2.2 The Assurance of Digital Assets (ADA) Framework*
The ADA framework is a structured approach for the assurance of digital assets, which was developed and matured on the case of an autonomous ferry [8]. This paper formally presents the ADA framework with CBD and SBT principles, for the ferry use case. The ADA framework (Fig. 4.1 in [8]) has two stages, presented in Sec. 2.2.2 & 2.2.3, and three central principles reflecting the assurance strategy: (1) *iterative* assurance, assuring over repeated cycles, to tackle complexity and continuously refocus attention, (2) *modular* assurance, by splitting into manageable modules, to tackle complexity and business collaboration, and (3) *continuous* assurance, to support continuous deployment during the whole system lifecycle.

*2.2.1 The ADA system model and responsibility model for control objectives*
The CBD framework from [3] (Sec. 2.1) models only generic system responsibility structures, but specific modelling support for system and risk identification, and transition of responsibility from a subsystem to a system level in control systems is needed. The System-Theoretic Process Analysis (STPA) control structure [9] is well-proven for control systems (Fig. 4.6 & 3.2 in [10]). However, it does not immediately fit with the CBD framework. Therefore, adapted system and responsibility models, shown in **Fig. 1**, have been developed in the ADA framework.

**Fig. 1** (left) shows the *system model* for a control system, which uses four component types: *Decision type*, generating decisions that will lead to system behaviour (STPA controller), *SITAW type*, generating beliefs about external/internal states or conditions of the system (STPA process model), *Action type*, converting decisions into actions, and *Resource type*, providing resources or support to any other component type. Risk sources (RS) specific to the component types, numbers 1-4 in **Fig. 1** (left) are adopted from STPA types of causal factors [9]. The risk-informed system model helps guide the system and risk identification approach in the ADA framework (Sec. 2.2.4). Risk sources highlight specific dependencies between components that can be handled by specifying CBD contracts. STPA Step 3 & 4 [11] are used for risks of inadequate decisions (STPA unsafe control actions) and causal factors from RS 1 & 3. If the Action component is implemented by a control system, RS 4 is covered in the same way, via an iteration on its subsystem. Other cases, e.g. for RS 4 or for RS 2, are not discussed further in this paper.

The *responsibility model,* **Fig. 1** (right), shows the allocation of responsibility (R) within a control system and guides the specification of CBD contracts, showing how responsibility is delegated. The Decision component takes the overall system responsibility ($R_{SYS}$), like leaders in organizations. This does not mean it performs all tasks, as the responsibility is further split and delegated: (i) the Decision component makes the proper decisions (Decision $R_2$), (ii) the SITAW component provides the necessary beliefs (SITAW $R_1$), (iii) the Action component converts the decision into action and system behaviour (Action $R_1$). The SITAW or Action components will always have a limited capability due to the level of accuracy of beliefs or actions. Therefore, they perform their delegated tasks to a certain quality level, reflected in their responsibilities (SITAW $R_1$ and Action $R_1$). For the Decision component to take the overall responsibility, it must account for these limited capabilities in its decision, and not expect more. Thus, there is a clear consistency and structure in how responsibilities are delegated. This is used in the ADA framework's CBD contract structure of control systems, see Section 2.3.

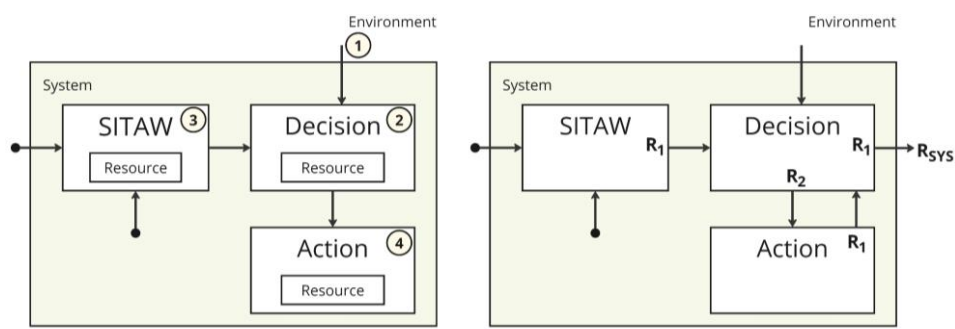

**Figure 1.** ADA system model (left) with risk sources RS 1-4 (circled numbers), and ADA responsibility model (right) with responsibilities ($R_i$) per component.

*2.2.2 ADA Stage 1 – Profile the digital asset*
ADA Stage 1 helps build a *profile of the digital asset* to support iterative assurance. Its focus is on an operational, overarching ecosystem around the asset, with three identification goals: (1) the digital asset (conceptually), (2) the context of use, and (3) the asset requirements to meet safety, efficiency or stakeholder need(s). Stage 1 is organized in steps, normally executed iteratively.
**Fig. 2** shows the different steps, and the relation to ADA Stage 2 and ADA System Identification.
**Step 1 - Understand the asset and its use** by describing the asset at a conceptual level, the idea, and expected use.
**Step 2 - Identify the context of use** to understand the operational environment and the extended ecosystem properly. CBD contracts can document the main asset components, entities, and any relevant information of the operational design domain (ODD).
**Step 3 - Analyse the stakeholders, their needs, and concerns** to provide trust that any relevant stakeholder need is sufficiently fulfilled, or concerns are ungrounded.
**Step 4 - Identify asset-related goals, losses, hazards, and functional requirements** based on stakeholder needs and concerns. The requirements are documented in asset CBD contracts, which make up the specific assurance goals.

ADA Stage 1 uses some critical methods to support a more complete set of functional requirements: **Context mapping** is used in Step 2. It is an established and formal modelling approach to create system context diagrams [12] that map *entities* (incl. objects, stakeholders and conditions) and their *relations* in a certain context. This information is necessary for the quality of Step 3 & 4. From the context mapping, a complete list of stakeholders can be extracted. A **stakeholder analysis** is then used in Step 3, to understand any stakeholder need, concern, risk, or trust need, in the ecosystem around the deployment [13]. The stakeholder analysis merges key aspects of "*jobs to be done*" [14] and how to deliver value using *pains* and *gains* analysis [15] and can generate key assurance goals and requirements in Step 4. **Loss and hazard identification** is used in Step 4 to identify hazardous situations and behaviours that could lead to loss. This is an improvement on *STPA Step 1* [11], focusing on any relevant operational step or situation, while guided by a list of harm source types. This can lead to more complete and specific functional requirements.

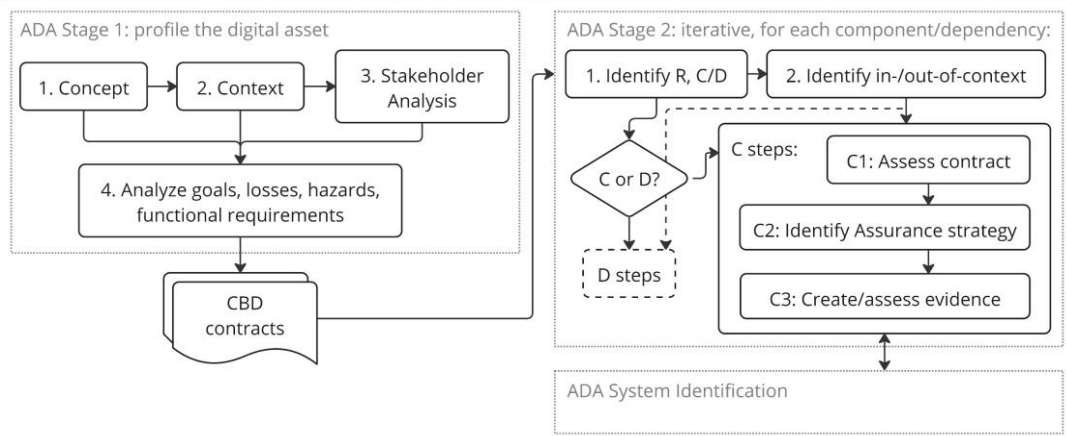

**Figure 2.** ADA stages and steps.

*2.2.3 ADA Stage 2 – Substantiate claims reflecting stakeholder needs*
ADA Stage 2 changes perspective by offering modular and iterative assurance of components at any system level, relevant for stakeholder needs. Stage 2 offers core assurance guidance based on the CBD method: *Component assurance* to assure that a component contract is sufficiently complete, and that the implementation sufficiently satisfies it; and *Dependency assurance* to assure the consistency (refinement) in the component dependencies. ADA Stage 2 consists of:
**Step 1 – Identify responsibility and component, or dependency, needing assurance**: select which *responsibility* (requirement) and which *component* (C steps) or *dependency* (D steps) between components to focus on.
**Step 2 – Identify the assurance context**: assess if assurance happens in (a) a well-defined *in-context* case with a Stage 1 profile, or in (b) a less-defined and *out-of-context* case where profile assumptions must be made? We adopt the same definition of *out-of-context* as ISO 26262 [16].
*C* **steps– Assurance of a component**. To support Step C1 and insight evidence of Step C2, the ADA system identification approach, further explained in Section 2.2.4, is used.
**C1 – Assess component contract**: assess **(1)** the contract guarantee (G) and how it fits the given responsibility and dependency from other components/entities in the environment, and **(2)** the contract assumptions (A), reflecting component-external dependencies to satisfy G.
**C2 – Decide a component assurance strategy**: make a strategy from three types of evidence:
**(1)** *Observations* of the component's behaviour, from any kind of testing, e.g. trials, simulations, or operational data. **(2)** *Insights* and information from the component's implementation, internal function, or processing. Note that this is technology dependent. For complex systems, the mapping and assurance of the subcomponents and their dependencies in a recursive and iterative way is important. **(3)** *Circumstantial* evidence about the development or responsible organisation's knowledge, capabilities, or ability to succeed with the component implementation. The evidence types relate to the *extrinsic* (*observations*) and *intrinsic* (*insights* and *circumstantial*) *causes-of-trust* concept in [17]. *Observation* and preferably *insight* evidence should be specified/ described compatible with the contract *A* & *G* terms, reflecting how certain component behaviours, under specific states of the component environment and dependencies (*A*), match the contract *G*. This can help formalize the strength, relevance, and coverage of the strategy and the assessment in Step C3.
**C3 – Create and assess evidence**: produce/collect evidence and assess its strength, relevance, or coverage towards the component contract.
*D* **steps – Assurance of a dependency between components** to build confidence in that a component does not *need* (*assume*) more than what is *provided* (*guaranteed*). This assessment is guided by the criticality and risk in the dependency, and the level of confidence in what is *provided*. These steps are mainly specification tasks for the *needs* and component assurance tasks for the *providing* components, used by subsystem assurance strategies in Step C2. Given space constraints, this paper does not showcase the steps in D and further details are left out.

*2.2.4 ADA System Identification using decomposition and a suitable risk model*
The CBD and ADA frameworks are supported by system identification and decomposition to integrate and assure complex systems. This is made for any component type or assurance context (Step C1 & C2 of ADA Stage 2, Section 2.2.3). This paper only explores a control system, using the system and responsibility models from Section 2.1.
**Step 1 – Identify the component and responsibility in focus for the system identification**: one component and one responsibility (guarantee) at a time.
**Step 2 – Identify the component subsystem type**: only assurance guidance for a control system is discussed here.
**Step 3 – Identify the responsible subcomponent**: identify which subcomponent takes overall responsibility. For a control system type, this should point to a *Decision component*, inheriting the responsibility and guarantee, added to the *R & G* fields in its contract (previously explained in **Tab. 1**). In case the overall responsibility is shared, further identification needs to be split for each identified Decision component.
*Steps C – Guidance for system identification of a control component type*:
**Step C1 – Identify subcomponents of control system type**: control actions that the Decision component generates can be identified. This reveals the interface and split of responsibility between the Decision and Action component, and contract *R*, *function*, *output,* and *input* fields can be documented.
*Steps C2 – Do an interaction analysis around a decision component*:
**Step C2.1 – Identify scenarios of inadequate or unsafe control**: use *STPA Step 3* [11] on the identified control actions generated by the Decision component, under the relevant operational hazardous context (identified in ADA Stage 1). The result is a list of inadequate control scenarios, handled in Step C2.2.
**Step C2.2 – Identify causal factors for scenarios of inadequate or unsafe control**: employ *STPA Step 4* [11] on each inadequate control scenario. This produces causal factors belonging to the four RS types shown in **Fig. 1** (left): (1) control inputs to the Decision component, (2) implementation and capability of the Decision component, (3) SITAW inputs to the Decision component, and (4) Action component capability.
**Step C2.3 – Identify component requirements from the causal factor scenarios**: assess the risk of each scenario, before specifying mitigation through the contracts: **Causal Factor Type 1** means a dependency on higher level control input, which can be added as a Decision component contract *A*. **Causal Factor Type 2** emerges from within and is handled in a separate system identification of the Decision component. **Causal Factor Type 3** relates to the Decision component dependencies on a *SITAW* component, can be added as requirement pairs (A, G) of Decision contract *A* and SITAW contract *G* respectively. **Causal Factor Type 4** relates to the Action component capability and the underlying causes must be revealed in a separate system identification effort.

The responsibility split discussed in Section 2.1, shown in **Fig. 1** (right), can guide specifying the guaranteed Action capability and added to the *R*, *function,* and *G* fields of the Action contract, and to the *A* field of the Decision contract.

## 3. Applying ADA on the ferry use case

We present below the practical use of the ADA framework on the ferry use case, with a focus on collision avoidance during the ferry's transit phase. The key outcomes are (1) contracts for identified components, and (2) modular assurance cases with evidence needs, for components we focus on, such as the ferry itself and the collision avoidance system.

*3.1 Results from ADA Stage 1 (following the steps in Section 2.2.2)*
**Step 1:** The ferry system concept and principle of operation, such as collision avoidance in transit, are identified. **Step 2:** The operational context is identified including stakeholders, types of expected vessels, objects, or obstacles relevant for collision avoidance during transit. **Step 3**: The specific needs or concerns of the identified stakeholders are analysed, including what is relevant for the ferry's transit phase. **Step 4:** A loss and hazard analysis is conducted, followed by identifying functional requirements. Combining this information with information from Step 1 and 3, a ferry contract for collision avoidance is specified with its *R*, function, and *G* fields. The ferry shall keep a minimum separation distance from any obstacle by adjusting its speed on its predefined crossing path. Ferry assumptions about a predefined path, obstacle properties and the operational environment are also added to its contract.

*3.2 Results from ADA Stage 2 – Iteration 1 (following the steps in Section 2.2.3)*
**Step 1:** The focus is on assurance of the ferry for collision avoidance during transit, captured in a CBD contract from ADA Stage 1, Step 4. **Step 2:** The assurance context is in-context, with full knowledge from the profile in Stage 1. **Step C1:** The ferry shall satisfy the CBD contract from Stage 1, Step 4. Assess that (1) the guarantee can/should be fully specified based on the analysis conducted in Stage 1, Step 4, and any risk and dependency that were identified. Then assess (2) the assumptions based on the knowledge from the ferry and transit ODD Stage 1 contracts. Further assumption knowledge is not available until the ferry is further analysed or assured. **Step C2:** Find an assurance strategy considering (1) observation, and (2) insight evidence. Observing only real ferry operations does not provide enough evidence and there are also many stakeholders that require more transparency and insights, especially system/component providers who must tackle system integration challenges. **Observation** evidence: (1) Real testing or operation of the employed ferry, e.g. field trials or capturing data from real operations, (2) Model-based end-to-end simulation of the ferry operation. **Insight** evidence: (1) System identification and risk analysis: to increase the transparency, knowledge, and confidence in its robustness, and (2) Modular assurance of the ferry subsystem, its key components and integration: to gather knowledge, justify confidence and provide integration help to system providers. The assurance strategy requires the system identification effort to identify these key components and their dependencies. **Step C3:** The observation evidence clearly specifies real testing and SBT activities. The insight evidence points to a ferry system identification effort (Sec. 3.3), and a modular assurance effort, presented here as assurance of the Motion Planning Control System (MPCS) component, detailed in Section 3.4.

*3.3 Results from the ADA system identification (Section 2.2.4) of the ferry component*
**Step 1**: Iteration 1 of ADA Stage 2 requires a system identification of the ferry for collision avoidance during transit. **Step 2:** The ferry collision avoidance is of a control type. **Step 3**: Based on **Fig. 1**, MPCS has the overall responsibility, which is added to its *R & G* contract fields, as shown in **Fig. 3**. **Step C1:** It is found that MPCS issues setpoints to the dynamic positioning (DP) component to manoeuvre the ferry. The MPCS's *function*, *R & G* contract fields, and the DP's *R*, *function*, *input* contract fields can be specified. **Step C2.1**: From analysing the collision scenarios, the following unsafe control scenarios were identified: MPCS not providing setpoints or ordering DP a too high or too low speed, a wrong position or heading. **Step C2.2 & C2.3**: Causal factor types help to identify MPCS' dependencies (*A*) to other components' contract *R & G* fields, see **Fig. 3** and **Tab. 2**. **Causal Factor Type 1:** MPCS must assume (*A*) that the ferry responsible person/system correctly configures the ferry's preset path, safe speed, and safety distance to obstacles. **Causal Factor Type 3:** Two sources of uncertainty are found in the estimated and predicted necessary state data about (1) obstacles and (2) the ferry, which MPCS must assume (*A*). This also identifies (1) obstacle and (2) ferry state data of the SITAW component (**Tab. 2**). **Causal Factor Type 4:** Points to the DP and Ferry automation Action capability to manoeuvre accurately and failure free according to the setpoint. The DP's *R & G* must be updated with this capability promise that MPCS assumes (*A*).
Note for ADA Stage 2, we focused on the MPCS component of the ferry, resulting in **Fig. 3** and **Tab. 2**. Further A/G for all other components are to be determined in further iterations.

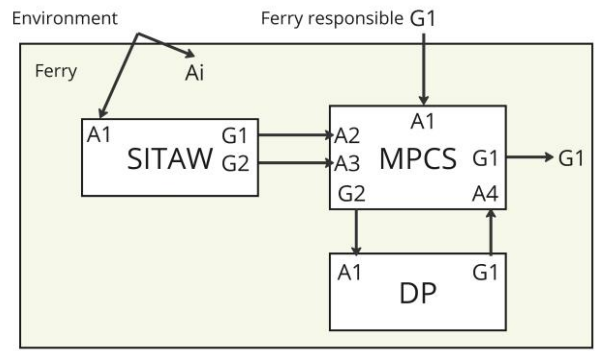

**Figure 3.** ADA contract with Assumptions (A) and Guarantees (G) for each system component, assessed from the perspective of the MPCS component. Table 2 describes all A and G.

*3.4 Results from ADA Stage 2 – Iteration 2 - MPCS*
We executed a **second iteration** of ADA Stage 2. From the first iteration, modular assurance of the Ferry subsystem for collision avoidance in transit is needed. The MPCS component is now selected to satisfy the CBD contract (**Fig. 3** & **Tab. 2**), and we can assess (Step 3) that the MPCS contract is well-defined from the previous system identification effort (Section 3.3). When deciding an assurance strategy (Step C2) at this level, observation evidence is most relevant for our paper, thus insight evidence is not discussed further. The key evidence is via SBT of the MPCS collision avoidance functionality (Step C3), see Section 4.

**4. Towards combining CBD and SBT in assurance**

The outcomes of the ADA iterations for the assurance of an autonomous ferry, focusing on collision avoidance during the ferry's transit phase, include: a specific observation evidence need and a CBD contract for the MPCS collision avoidance functionality. Via SBT, we can observe component behaviour to provide evidence of CBD contract satisfaction and that the CBD contract sufficiently represents the behaviour of the component. This shows the iterative nature of modular assurance and the CBD concept of refinement: A component implementation shall *refine* its CBD contract, which can be shown by generating evidence in an iterative and consistent manner, guided by CBD. If it does not *refine* its contract, the implementation must be improved and/or the contract relaxed, the system integration adjusted, and assurance iterated.
   The key task of SBT is thus to provide sufficient and trustworthy evidence to confidently compare a CBD contract and its component implementation, to confirm or falsify refinement. If we consider the contract for MPCS.G2 (**Tab. 2** & **Fig. 3**), we require that the MPCS component can provide the next setpoint for the DP component, to keep the ferry in a safe state under all assumptions A1 - A4. To demonstrate that the MPCS component satisfies its CBD contract, it is necessary to build confidence in, and a valid justification for, the SBT strategy. This depends on three central issues: (1) that the SBT scope sufficiently covers the system's (here: MPCS and its environment) behavioural space needed for conclusions, (2) that the observations (test results) are compatible with the component *acceptance criteria*, and (3) that the simulation model sufficiently represents the environment of the component. Based on this and utilizing CBD, we can express three key research questions and tasks:

*1. How can a CBD contract be used to specify and assess an **SBT scope**?* The challenge is to produce a test plan with a clear reference to the component dependencies and contract assumptions, but where the basic task is to "*cover the behavioural space for conclusions*". The test scope is defined by the combination of the assurance task and the contract, to validate that the component satisfies the CBD contract. The test scope needs to cover the space of assumptions, guarantees and possible system behaviours. The two objectives of the test are to observe component behaviour to make conclusions about contract satisfaction, and whether the contract sufficiently represents the behaviour of the component. Based on the contract, one can identify suitable simulation environments, conditions, and parameters that the test cases shall explore to fulfil the observation task (see point 3).

*2. How can a CBD contract be used to specify SBT observation **evaluations**?* This involves formalising the test evaluation criteria. It should be able to assess if and to which degree the behaviour satisfies the guarantee or not. The criteria should therefore include specific indicators that are directly linked to the guarantee, i.e. quantitative indicators, such as performance metrics. One of the requirements for MPCS.G1 is to ensure that the provided setpoint maintains a sufficient safe distance. As we can quantify safe distance as a metric, we can use it as one of our evaluation criteria to verify that the guarantee is satisfied. This also points back to that the guarantee (MPCS.G1) must be sufficiently exact and quantitative. Note that for cases where there is a large observation space, e.g. in testing collision avoidance, there is a need for efficient automated testing. Thus, the evaluation metrics should support this.

**Table 2.** Contract specifications after assurance and system identification iterations, ADA Stage 2, focusing on the MPCS component. Definitions: "Vessel/obstacle state data" is position, speed, heading, course; "w.m.a." is "within a minimum (agreed) accuracy". The A/G relationships are shown in Fig. 3.

| Component | Assumptions & Guarantees |
|---|---|
| Ferry | **Ai**: Assumptions aggregated from ferry components about the environment. Will be identified in further iterations. |
| Ferry | **G1**: Guarantees to keep a safe minimum distance to obstacles. |
| MPCS | **A1**: Assumes it is configured with a valid route and safe minimum distance.<br>**A2**: Assumes it receives all estimated/predicted obstacle state data/dimensions w.m.a.<br>**A3**: Assumes it receives all vessel state data w.m.a.<br>**A4**: Assumes that "DP" manoeuvres ferry into desired state w.m.a. |
| MPCS | **G1**: Inherits responsibility for Ferry G1.<br>**G2**: Guarantees to provide next setpoint to keep ferry in a safe state, accounting for SITAW and "DP" accuracies. |
| SITAW | **A1**: Assumes properties and behaviours of obstacles relevant for the location. |
| SITAW | **G1**: Guarantees to provide estimated/predicted obstacle state data/dimensions, w.m.a.<br>**G2**: Guarantees to provide all ferry (ownship) vessel state data w.m.a. |
| "DP" | **A1**: Assumes it receives a desired setpoint (vessel state) to bring the ferry into. |
| "DP" | **G1**: Guarantees to manoeuvre the ferry into desired vessel state w.m.a. |
| Ferry resp. | **G1**: Guarantees the ferry is configured with a valid route and safe minimum distance. |

*3. How can a CBD contract be used to specify and assess the **simulation model** used for the SBT scope?* To ensure that the simulation results are valid and justify the observation evidence strategy, the simulation model should sufficiently represent the component environment. The contract assumptions specify the component dependencies to its environment and are key candidates for the simulation model requirements. These can be used to build and assure a simulation model. In SIL context, each simulation model can be treated as a black box and tested individually, to ensure it behaves as expected when integrated within the whole system. However, each simulation model of any system component that is not directly under test, e.g. SITAW, still needs to meet the info and accuracy requirements given by e.g. MPCS.A1.

The exploration of these three research questions is the topic of ongoing research, to determine how the CBD contract for a system can be used to define the 1) test scope, 2) evaluation criteria, and 3) simulator requirements. This can then be used to define the test strategy for the system via formalisation [18] and explorative *design of experiments* (DoE) methods [19]. Using DoE in automated testing allows to adaptively optimize for multiple criteria, e.g. evaluation criteria stemming from the CBD contract, and the exploration-exploitation trade-off within the parameter search spaces. This will ensure the necessary number of tests is done to obtain the evidence for the assurance case of the autonomous system.

## 5. Conclusion

This paper proposed a new holistic strategy that integrates contract-based design (CBD) and simulation-based testing (SBT) to enhance the assurance processes for ASVs. The ADA framework provides a structured approach that utilizes CBD for the iterative, modular, and continuous assurance of digital assets. CBD, by enabling modularity, gives a robust foundation for decomposing complex systems into manageable components, each with defined contracts specifying their responsibilities and dependencies. Building on this, we explain how SBT can be used to provide evidence for verifying CBD contracts in complex operational scenarios, to build the necessary evidence for an assurance case.

This paper showcased the results of applying the ADA framework to the assurance case of an autonomous ferry, focusing on the collision avoidance functionality in the MPCS component, during the vessel's transit phase. When applying the ADA framework to assurance of the MPCS component, CBD ensures that the component functions within its defined parameters, well integrated in the ferry system. Consequently, this assurance strategy not only facilitates rigorous verification processes but also allows for detailed exploration and validation of each contract. The proposed ADA framework is scalable and adaptable and can be applied to a wide range of autonomous, complex, or intelligent systems in various sectors.

Future work aims to demonstrate additional practical applications of the ADA framework and will attempt to answer the key identified research questions on the application of CBD towards SBT of autonomous systems.

**Acknowledgements**

This works was funded in part by TRUSST: NFR MAROFF-2 project nr 313921 (Assuring Trustworthy, Safe and Sustainable Transport for All), and SIMPLEX: NFR MAROFF-2 project nr 341045 (The role of SIMulation in assurance of intelligent and comPLEX systems).